\documentclass[jkps,preprint,fleqn,showkeys]{revtex4}
\usepackage{graphicx}
\usepackage{amssymb}
\usepackage{epstopdf}
\usepackage{epsfig,graphicx}
\usepackage[english]{babel}
\usepackage{amsfonts}
\usepackage{amsmath}
\usepackage{latexsym}
\usepackage{graphics,bm}
\usepackage{dcolumn}
\usepackage{natbib}
\usepackage{bm}
\usepackage{rotating}
\begin{document}
\setcounter{page}{0}
\title[]{Role of Dissipation on the Stability of a Parametrically Driven Quantum Harmonic Oscillator}
\author{Subhasish \surname{Chaki}}
\affiliation{Department of Chemistry, Indian Institute of Technology, Bombay, Powai, Mumbai 400076, India}
\author{Aranya bhuti \surname{Bhattacherjee}}
\email{aranyabhuti@hyderabad.bits-pilani.ac.in}
\affiliation{Department of Physics, Birla Institute of Technology and Science, Pilani, Hyderabad Campus, Hyderabad 500078, India}

%%%%%%%%%%%%%%%%%%%%%%%%%%%%%%%%%%%%%%%%%%%%%%%%%
%               TITLE 
%%%%%%%%%%%%%%%%%%%%%%%%%%%%%%%%%%%%%%%%%%%%%%%%%

%\date{\today}

\begin{abstract}

\noindent \textbf{Abstract:} We study the dissipative dynamics of a single quantum harmonic oscillator subjected to a parametric driving with in an effective Hamiltonian approach. Using Liouville–von Neumann approach, we show that the time evolution of a parametrically driven dissipative quantum oscillator has a strong connection with the classical damped Mathieu equation. Based on the numerical analysis of the Monodromy matrix, we demonstrate that the dynamical instability generated by the parametric driving are reduced by the effect of dissipation. Further, we obtain a closed relationship between the localization of the Wigner function and the stability of the damped Mathieu equation. 
\end{abstract}


\maketitle

\section{Introduction}
\noindent The quantum harmonic oscillator is one of the simplest but important models of microscopic and macroscopic systems in many branches of physics such as quantum optics \cite{pedrosa2005exact}, ion-traps \cite{leibfried2003quantum}, quantum dots \cite{englund2005controlling},  quantum information \cite{schleich2007elements}, and superconductivity \cite{wallraff2004strong}. In real physical situations,  damping is a key issue which makes it considerably difficult to analyze. However, there are two main approaches to understand dissipation in quantum systems at a fundamental level. The first one is to put a system in a bath of many harmonic oscillators where the bath oscillators interact with the system through certain coupling \cite{zwanzig1973nonlinear}. The system and bath oscillators together make a conservative system and the rule of quantization is applied to that conservative system. Then, the dissipation in the system is obtained by eliminating the bath degrees of freedom. Thus, the quantum mechanical fluctuation-dissipation relation is established due to the coupling where the bath oscillators exert force on the system as fluctuation and the system gives back its stored energy to the bath oscillators through the coupling as dissipation. The second approach is based on the formulation of an effective Hamiltonian  which basically represents the classical model of quantum damped harmonic oscillator (QDHO). In the 1940s, Caldirola \cite{caldirola1941forze} and Kanai \cite{kanai1948quantization} independently introduced the QDHO  model Hamiltonian known as CK Hamiltonian  which is an one-dimensional system with an exponentially increasing mass. The advantage of CK Hamiltonian is that it is an open system whose parameters such as mass, frequency are time-dependent. One can easily derive the equation of motion of a harmonic oscillator subjected to frictional force  from CK Hamiltonian. But, it should be noted that there is no external force involved in the system. At long time, the quantization of CK Hamiltonian gives rise of the violation of uncertainty principle because the system is losing energy exponentially with time \cite{dekker1981classical}. To overcome this unphysical violation of uncertainty principle, an external force can be added in a certain way to the CK Hamiltonian  \cite{dodonov1979coherent}. Thus, the CK Hamiltonian which only takes into account the damping is not sufficient  to describe the quantum dissipative system. Related to CK model, a detailed theory of Brownian motion of a quantum oscillator was developed in 1971 \cite{agarwal1971}. The author had employed the phase-space distribution functions obtained from the density operator via certain rules of mapping.The resulting equation for the reduced
phase-space distribution function is found to be of the Fokker-Planck type. It was also shown that the system reaches equilibrium as $t \rightarrow \infty$. More recently, using nonperturbative approach, dissipative dynamics of a harmonic oscillator was studied \cite{jishad2009}.In the limit of vanishing oscillator frequency of the system, the authors recover the results of the free Brownian particle.
\\
\\
An immediate interesting study would be to understand the role of dissipation for CK Hamiltonian in the presence of parametric driving. In general, a system is said parametrically driven if one, or more,of its parameters is varied periodically in time and the phenomenon of parametric resonance can also occur at multiple frequencies. Under parametric resonance of a harmonic oscillator, stable points of the undriven system become unstable and vice-versa  for specific values of the period of the parameter variation. Parametric resonance  occurs in a wide variety of systems such as classical oscillators, nonlinear optics, system governed by non-linear Schr$\ddot{\textrm{o}}$dinger equations and in Hamiltonian chaotic systems.  A cigar-shaped Bose-Einstein condensate (BEC) in a harmonic trap exhibits parametric excitations in the form of Faraday waves when the frequency of the trap is periodically modulated in time \cite{bhattacherjee2008faraday,verma2012oscillations}. These parametric excitations create instability in the system which tries to delocalize the BEC.  In another context, it has been shown that quantum phase transitions from normal to superradiant phase become prolific in a periodically driven Dicke model. However, the stability of those  phases are sensitive to the parameters of the time-periodic coupling \cite{bastidas2012nonequilibrium}. Usually, for parametrically driven systems, the dynamics are governed by a Mathieu equation and the usual technique for dealing with such systems is the Floquet formalism \cite{abramowitz1965handbook}. However, the role of dissipation in the context of parametrically driven quantum systems is still not much explored.
\\
\\
In this paper, we have studied the dynamics of a dissipative quantum harmonic oscillator for CK Hamiltonian whose parameters are periodically driven with time. The annihilation and creation operators associated with this dissipative parametrically driven quantum oscillator are time-dependent but satisfy the Liouville–von Neumann quantum canonical equation. We have shown that in the quantum domain, the dynamical behaviour of the system is fully determined by the solution of the corresponding classical damped Matheiu equation. Our main aim is to investigate the effect of dissipation on the stability of the system which is very essential in the context of parametrically driven systems. For that purpose, we have done the stability analysis by constructing the Monodromy matrix to check whether damping increases the stability or not for parametrically driven quantum systems. Furthermore, we have studied the  dynamical behaviour of the Wigner function for the same model and found that the behavior of Wigner functions are quite consistent with the stability profile of the system.
\section{Parametrically Driven CK Hamiltonian} 
\noindent We consider a CK Hamiltonian of a single quantum harmonic oscillator with time dependent parameters,
\begin{equation}
\begin{split}
\hat{H}_D (t)=\frac{1}{2}\left[e^{-\gamma t}\hat{P}^2+e^{\gamma t}\omega^2(t)\hat{Q}^2\right],
\label{eq:dissipative_hamiltonian}
\end{split}
\end{equation} 
\noindent where $\hat{P}$ and $\hat{Q}$ are dynamical variables and $\gamma$ is the damping constant. The condition, $\gamma >0$, ensures the depletion of energy from the system. For driving the system periodically, we assume $\omega^2 (t)=g-h \cos (\phi t)$, where $g$ is the static contribution. Here, $h$ and $\phi$ are the amplitude and frequency of time-modulated parts, respectively. From Eq. $(\ref{eq:dissipative_hamiltonian})$ one can easily derive the equation of motion describing a  parametrically driven damped classical harmonic oscillator
\begin{equation}
\begin{split}
\ddot{Q}+\gamma \dot{Q}+\omega^2(t) Q=0.
\label{eq:dissipative_classical}
\end{split}
\end{equation}
The system is driven out of equilibrium by a change of the time-dependent parameters of the model. To deal with such systems, we will follow the Liouville–von Neumann (LvN) quantum canonical formalism \cite{lewis1969exact}. We assume that the non-equilibrium dynamics of the system is still goverened by standard Schr$\ddot{o}$dinger equation \cite{flores2003nonequilibrium},
\begin{equation}
\begin{split}
i\frac{\partial|\psi(t)\rangle}{\partial t}=\hat{H}_D (t) |\psi(t)\rangle.
\label{eq:schrodinger}
\end{split}
\end{equation} 
\noindent In Lewis-Riesenfeld canonical method, the invariant operators $(\hat{O})$ should satisfy the time-dependent LvN equation
\begin{equation}
\begin{split}
\frac{d\hat{O}}{d t}=\frac{\partial \hat{O}}{\partial t}+i\left[\hat{H}_D (t),\hat{O}\right]=0.
\label{eq:Lewis-Riesenfeld}
\end{split}
\end{equation} 
\noindent The operator $\hat{O}$ satisfying Eq. (\ref{eq:Lewis-Riesenfeld}) can be used to construct the exact quantum states of the Schr$\ddot{o}$dinger equation (\ref{eq:schrodinger}) which is given by
\begin{equation}
\begin{split}
|\psi(t)\rangle=\sum_{n} c_n e^{i\theta_n t} |\lambda_n(t)\rangle,
\label{eq:schrodinger-states}
\end{split}
\end{equation} 
\noindent where $\hat{O}|\lambda_n(t)\rangle=\lambda_n |\lambda_n(t)\rangle$ and $\theta_n(t)=\int_0^t dt \langle \lambda_n(t)|i\frac{\partial }{\partial t}-\hat{H}_D(t)|\lambda_n(t)\rangle$.
\\
\\
Next, we introduce time dependent annihilation and creation operators, $\hat{a}_D(t)$ and $\hat{a}^{\dagger}_D (t)$, given by 
\begin{equation}
\begin{split}
\hat{a}_D (t)=A(t)\hat{Q}+B(t)\hat{P},
\label{eq:ladder_creation}
\end{split}
\end{equation} 
\noindent and 
\begin{equation}
\begin{split}
\hat{a}^{\dagger}_D (t)=A^{*}(t)\hat{Q}+B^{*}(t)\hat{P}.
\label{eq:ladder_annihilation}
\end{split}
\end{equation} 
\noindent The time dependence in $\hat{a}_D (t)$ and $\hat{a}^{\dagger}_D (t)$ comes from the complex coefficients $A(t)$ and $B(t)$. For all the times, the commutation relation between $\hat{a}_D (t)$ and $\hat{a}^{\dagger}_D (t)$ is 
\begin{equation}
\begin{split}
\left[\hat{a}_D (t),\hat{a}^{\dagger}_D (t)\right]=1.
\label{eq:ladder_commutation}
\end{split}
\end{equation} 
\noindent From Eq. (\ref{eq:ladder_commutation}), we obtain that 
\begin{equation}
\begin{split}
A(t)B^{*}(t)-B(t)A^{*}(t)=-i.
\label{eq:ladder_wronskian}
\end{split}
\end{equation} 
\noindent Using Eq. (\ref{eq:ladder_creation}) in Eq. (\ref{eq:Lewis-Riesenfeld}), we obtain 
\begin{equation}
\begin{split}
\dot{A}(t)\hat{Q}+\dot{B}(t)\hat{P}=-A(t) e^{-\gamma t}\hat{P}+\omega^2(t)e^{\gamma t}B(t)\hat{Q}.
\label{eq:ladder_dynamics}
\end{split}
\end{equation}
\noindent Thus, we get 
\begin{equation}
\begin{split}
\dot{B}(t)=-A(t) e^{-\gamma t}.
\label{eq:ladder_condition_1}
\end{split}
\end{equation} 
and
\begin{equation}
\begin{split}
\dot{A}(t)=B(t) e^{\gamma t} \omega^2(t).
\label{eq:ladder_condition_2}
\end{split}
\end{equation} 
\noindent By combining Eq. (\ref{eq:ladder_condition_1}) and Eq. (\ref{eq:ladder_condition_2}), we can write 
\begin{equation}
\begin{split}
\ddot{B}(t)+\gamma \dot{B}(t)+\omega^2(t)B(t)=0.
\label{eq:ladder_damped}
\end{split}
\end{equation} 
\noindent Hence from Eq. (\ref{eq:ladder_wronskian}), the Wronskian condition will be, 
\begin{equation}
\begin{split}
\left[\dot{B}^*(t) B(t)-B^*(t)\dot{B}(t)\right]=i e^{-\gamma t}.
\label{eq:time_dependent_wronskian}
\end{split}
\end{equation}
\noindent  Eq. (\ref{eq:ladder_damped}) is called damped Matheiu equation and hence it is not an area preserving ordinary differential equations (ODE). It should be noted that the above Eq. $(\ref{eq:ladder_damped})$ is structurally  same as that of Eq. $(\ref{eq:dissipative_classical})$, the only difference being that it is for the complex coefficient, $B(t)$.  However, the Wronskian condition (Eq. (\ref{eq:time_dependent_wronskian})) is survived by the the time-dependent nature of $B(t)$ from $e^{-\gamma t}$ term in Eq. (\ref{eq:time_dependent_wronskian}). This is similar to the observation that the positivity condition has to be satisfied, in order to maintain the uncertainty relation in the problem of quantum Brownian motion\cite{jishad2009}. The positivity condition is only satisfied above a certain breakdown temperature which depends directly on the oscillator's damping and inversely on its frequency \cite{jishad2009}.
\\
\\
\noindent The ground state of the time-dependent oscillator $(|0,t \rangle)$ must satisfy the condition, 
\begin{equation}
\begin{split}
\hat{a}_D |0,t \rangle =0.
\label{eq:ladder_ground}
\end{split}
\end{equation}
\noindent In the coordinate representation, the ground state can be written as $\psi_0(x,t)=\langle x|0,t \rangle$ and Eq. (\ref{eq:ladder_ground}) becomes
\begin{equation}
\begin{split}
\left(\dot{B} (t) e^{\gamma t} x + i B(t) \frac{\partial}{\partial x}\right) \psi_0(x,t) =0.
\label{eq:ladder_ground_x}
\end{split}
\end{equation}
\noindent The solution of the Eq. (\ref{eq:ladder_ground_x}) is given by, 
\begin{equation}
\begin{split}
\psi_0(x,t) = C_1 exp\left( \frac{i \dot{B}(t)e^{\gamma t}}{2 B(t)}x^2\right).
\label{eq:ground_solution}
\end{split}
\end{equation}
\noindent It should be noted that $B(t)$ is a complex quantity. Using the Eq. (\ref{eq:time_dependent_wronskian}), the normalized solution of Eq. (\ref{eq:ladder_ground_x}) will be 
\begin{equation}
\begin{split}
\psi_0(x,t) = \left(\frac{1}{2 \pi |B(t)|^2}\right)^\frac{1}{4} exp\left( \frac{i \dot{B}(t)e^{\gamma t}}{2 B(t)}x^2\right).
\label{eq:ground_solution_normalized}
\end{split}
\end{equation}
\section{Stability of Damped Matheiu Equation}
\begin{figure}
	\centering
	\begin{tabular}{cc}
		\includegraphics[width=0.4\textwidth]{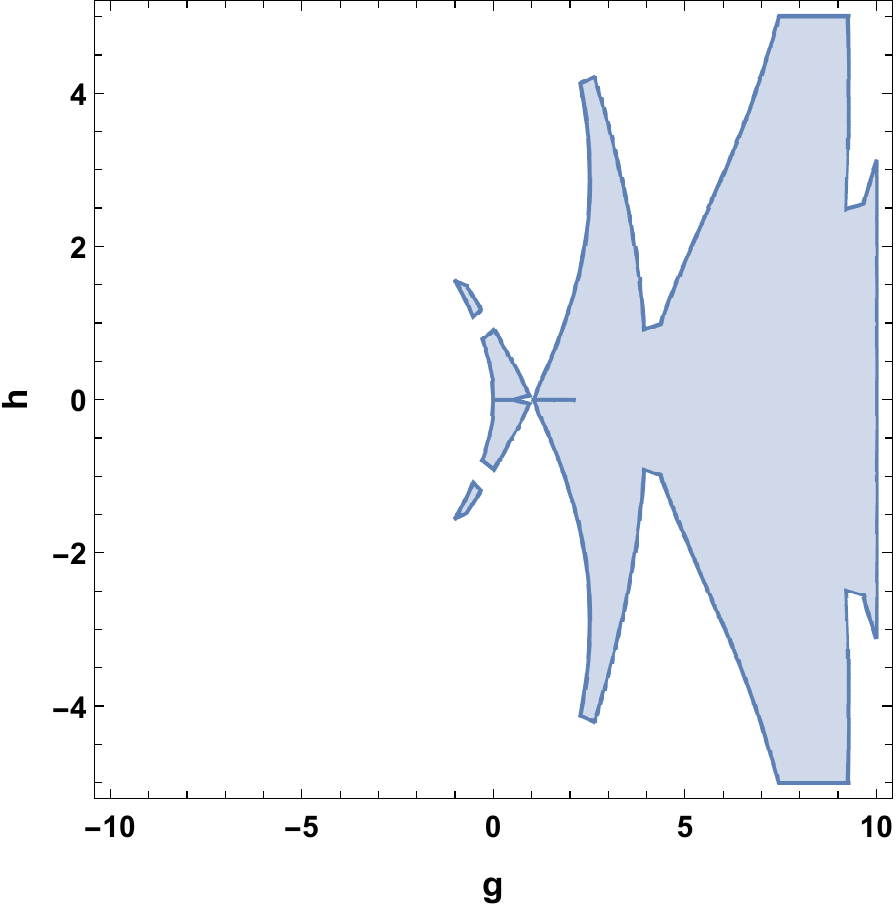} &
		\includegraphics[width=0.4\textwidth]{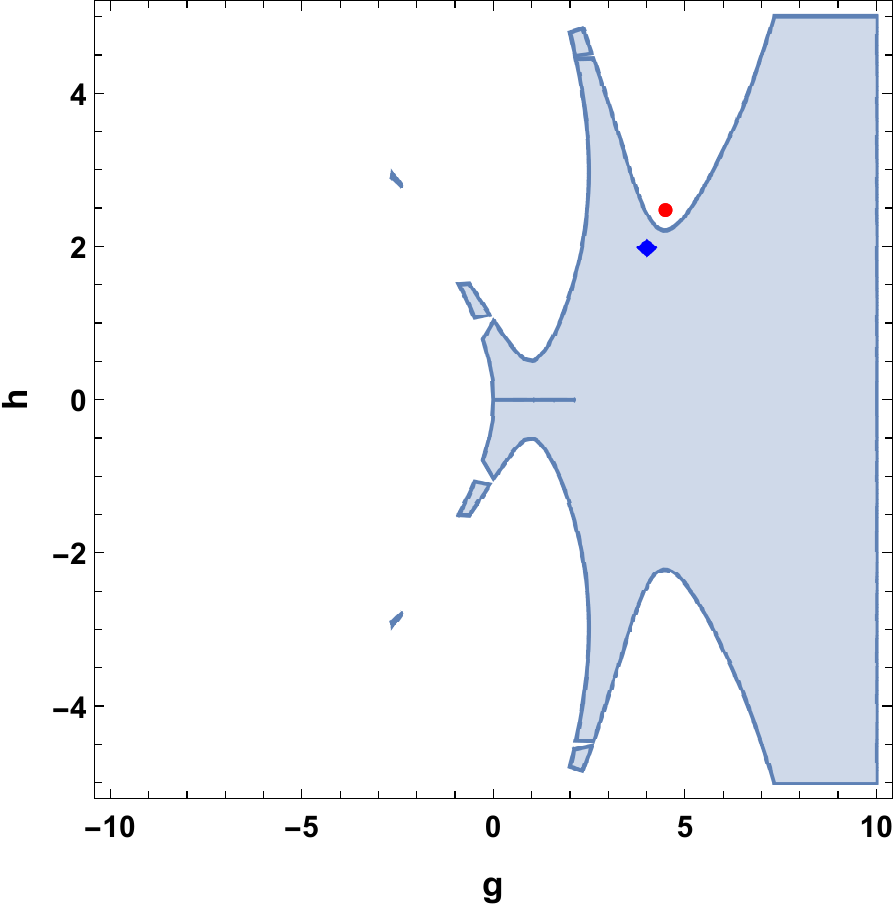} \\
		
		(a) & (b)  \\
		
	\end{tabular}
	\caption{Stability of the Mathieu’s equation for $\phi=2$: (a) without damping $(\gamma=0)$, and (b) with damping $(\gamma=0.5)$. }
\label{fig:Mathieu_stability}
\end{figure}

\begin{figure}
	\centering
	\begin{tabular}{cc}
		\includegraphics[width=0.4\textwidth]{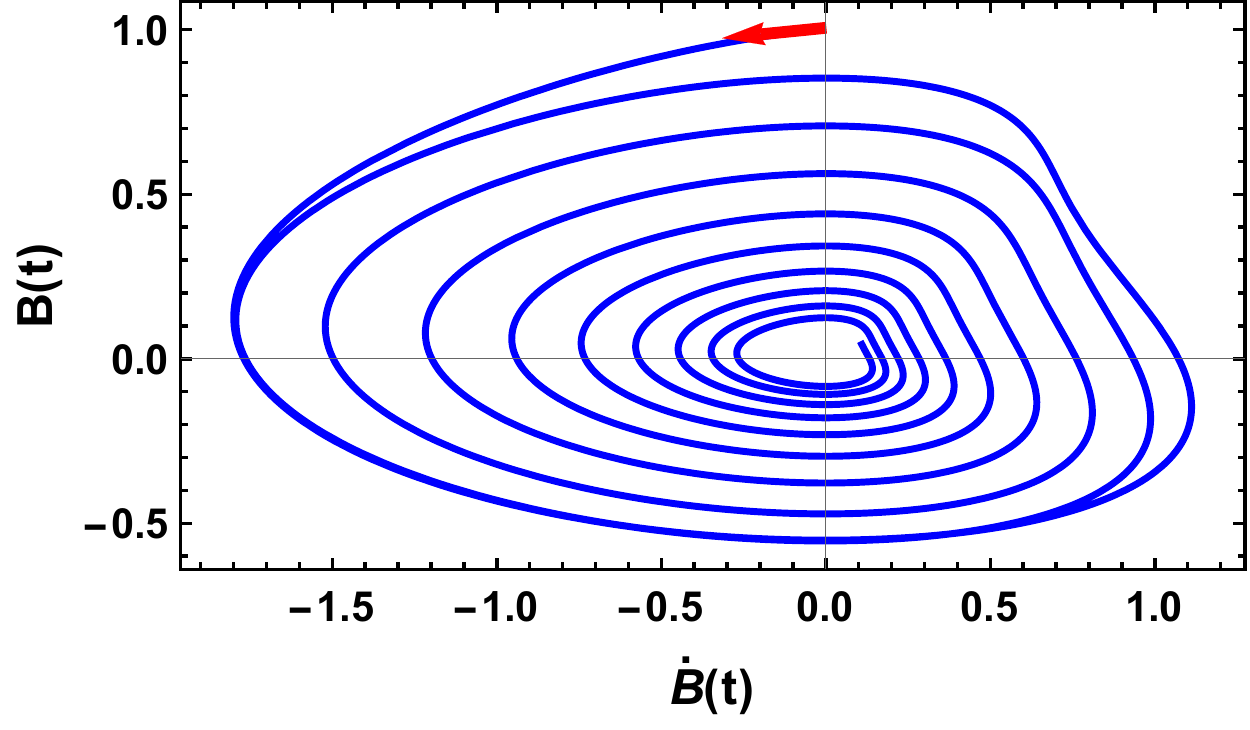} &
		\includegraphics[width=0.45\textwidth]{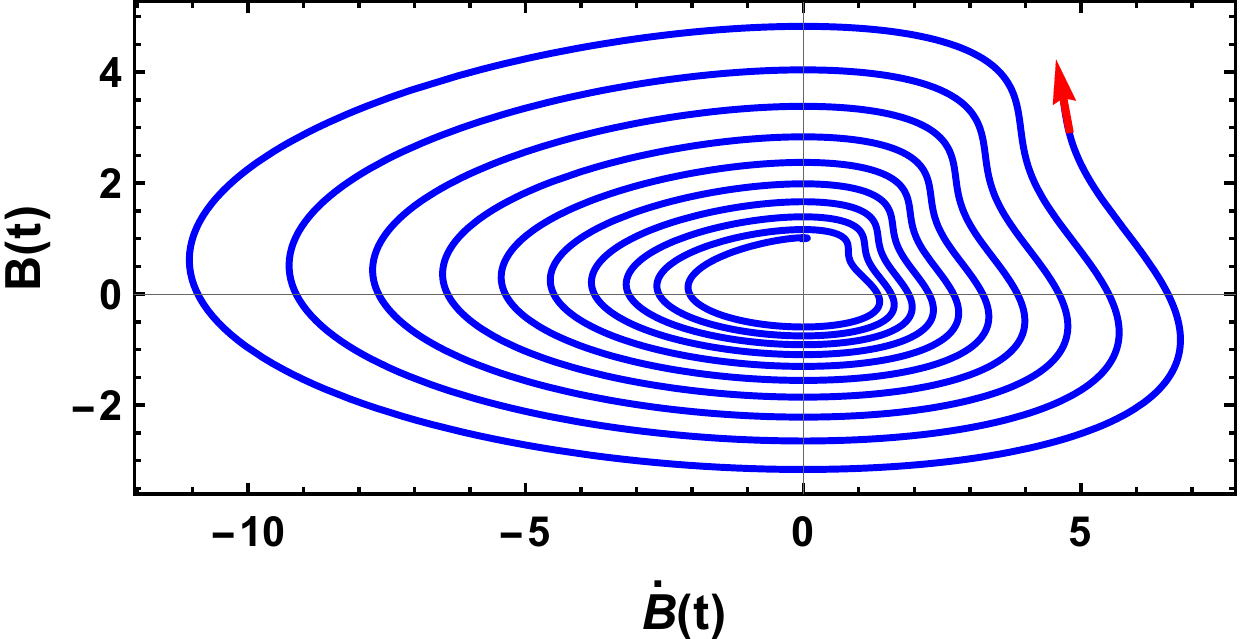} \\
		
		(a) & (b)  \\
		
	\end{tabular}
	\caption{Phase space trajectories of $B(t)$: (a) for stable blue diamond $(g=4.0, h=2.0)$, and (b) for the unstable red circle $(g=4.0, h=2.5)$. For both the plots (a) and (b), $\gamma=0.5$.}
\label{fig:Mathieu_damped}
\end{figure}

\noindent To explore the stability of the solutions of Eq. (\ref{eq:ladder_damped}), we apply the Floquet theorem to the damped Matheiu equation \cite{abramowitz1965handbook}. The two-dimensional first order differential equation associated with Eq. (\ref{eq:ladder_damped}) can be presented in a matrix form, 
\begin{equation}
\begin{split}
x^\prime =K(t) x,
\label{eq:first_order_matrix}
\end{split}
\end{equation}
\noindent where $x=\left[B(t),\dot{B}(t)\right]^{\dagger}$ and ` $\prime$ ' denotes the first order differentiation and $K(t)$ is a periodic function of time $(t)$ with periodicity $\tau=\frac{2\pi}{\phi}$. Using Floquet theorem, one can write the general solution of Eq. (\ref{eq:first_order_matrix}) as
\begin{equation}
\begin{split}
x(t) =e^{R t} P(t),
\label{eq:floquet}
\end{split}
\end{equation}
\noindent where $P(t+\tau)=P(t)$. Let $\{\lambda_i\}$ are the eigenvalues of $R$. The solution of Eq. (\ref{eq:first_order_matrix}) will be stable  if $\textrm{Re}(\lambda_i)<0$. Next, we will construct the Monodromy matrix $(M)$ to calculate $\{\lambda_i\}$ because  $\{\lambda_i\}$ are related to the eigenvalues of $M$, $\{\mu_i\}$    as $\mu_i=e^{\lambda_i \tau}$ \cite{insperger2003stability}. We will construct $M$ in the following.
\\
We can formally express $B(t)$ as
\begin{equation}
\begin{split}
B(t) =B^{(1)}(t)+B^{(2)}(t),
\label{eq:solution_matrix_B}
\end{split}
\end{equation}
\noindent and 
\begin{equation}
\begin{split}
\dot{B}(t) =\dot{B}^{(1)}(t)+\dot{B}^{(2)}(t).
\label{eq:solution_matrix_B_dot}
\end{split}
\end{equation}
\noindent Using Eq. (\ref{eq:solution_matrix_B}) and (\ref{eq:solution_matrix_B_dot}), we  construct the Monodromy matrix $(M)$ for the Eq. (\ref{eq:first_order_matrix}), 
\begin{equation}
\begin{split}
M=F^{-1}(0) F(\tau),
\label{eq:monodromy_matrix}
\end{split}
\end{equation}
\noindent where $F(\tau)=\begin{pmatrix}
  B^{(1)}(\tau) & B^{(2)}(\tau)\\ 
  \dot{B}^{(1)}(\tau) & \dot{B}^{(2)}(\tau) 
\end{pmatrix}$.  We have numerically constructed the matrix $M$ for the initial conditions,  $B(0)=1$ and $\dot{B}(0)=0$. and plotted the real part of $\{\lambda_i\}$ in Fig. $\ref{fig:Mathieu_stability}$  for $\phi=2\, (\textrm{or,}\, \tau=\pi)$.  The shaded band-like regions in the $g-h$ plane of Fig. $\ref{fig:Mathieu_stability}$  correspond to the stable zone whereas the white regions correspond to the unstable zone. With increasing the damping coefficient, more spreading of the band-like regions has been observed  in the $g-h$ plane of Fig. $\ref{fig:Mathieu_stability}$ (b) in comparison with Fig. $\ref{fig:Mathieu_stability}$ (a). This implies, damping increases the stability of the system \cite{nayfeh1980nonlinear}. A similar result was also found in the work of Agarwal \cite{agarwal1971} where the damped harmonic oscillator system was shown to reach equilibrium for large time.  For further investigation, we have chosen two different points from Fig. $\ref{fig:Mathieu_stability}$ (b): one is from the stable zone for which $g=4.0, h=2.0$ (blue diamond) and another is from the unstable zone for which $g=4.0, h=2.5$ (red circle). In Fig. $\ref{fig:Mathieu_damped}$, we have plotted the phase-space representation of the trajectories of $B(t)$. For stable blue diamond in Fig. $\ref{fig:Mathieu_stability}$ (b)  , $B(t)$ shows damped  oscillations in Fig. $\ref{fig:Mathieu_stability}$ (b) while for unstable red circle in $\ref{fig:Mathieu_stability}$ (b), $B(t)$ shows unbound oscillations in Fig. $\ref{fig:Mathieu_damped}$ (b).
\section{Wigner function Analysis}
\begin{figure}[h]
	\centering
	\begin{tabular}{ccc}
		\includegraphics[width=0.33\textwidth]{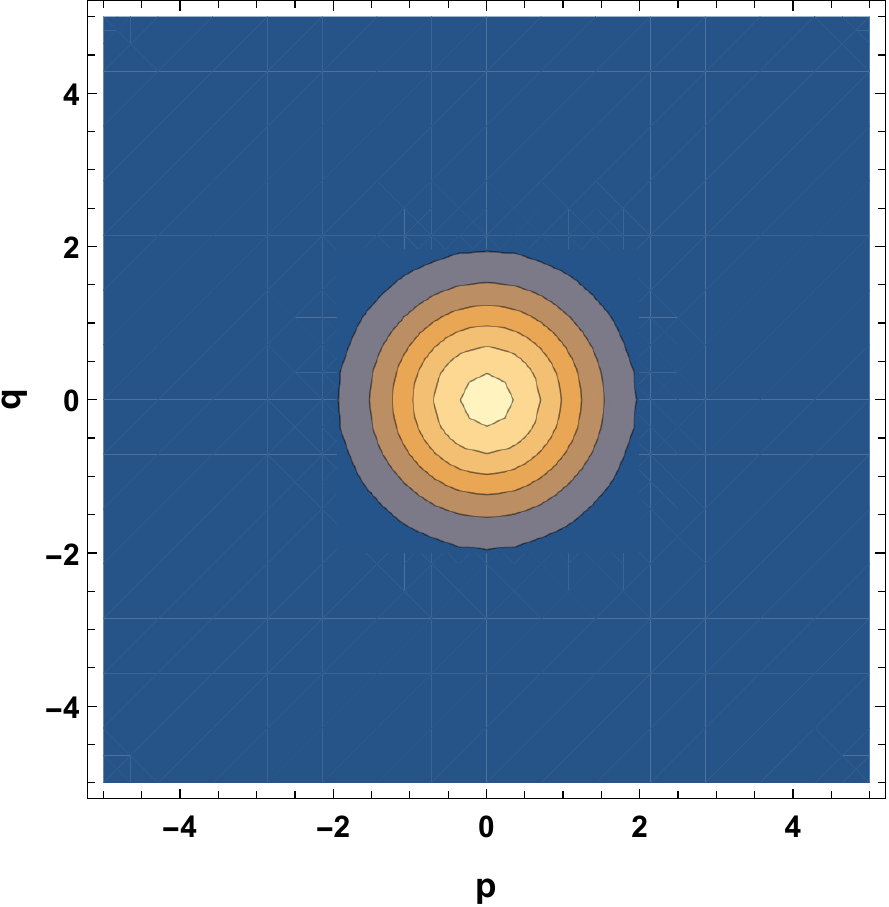} & 
		\includegraphics[width=0.33\textwidth]{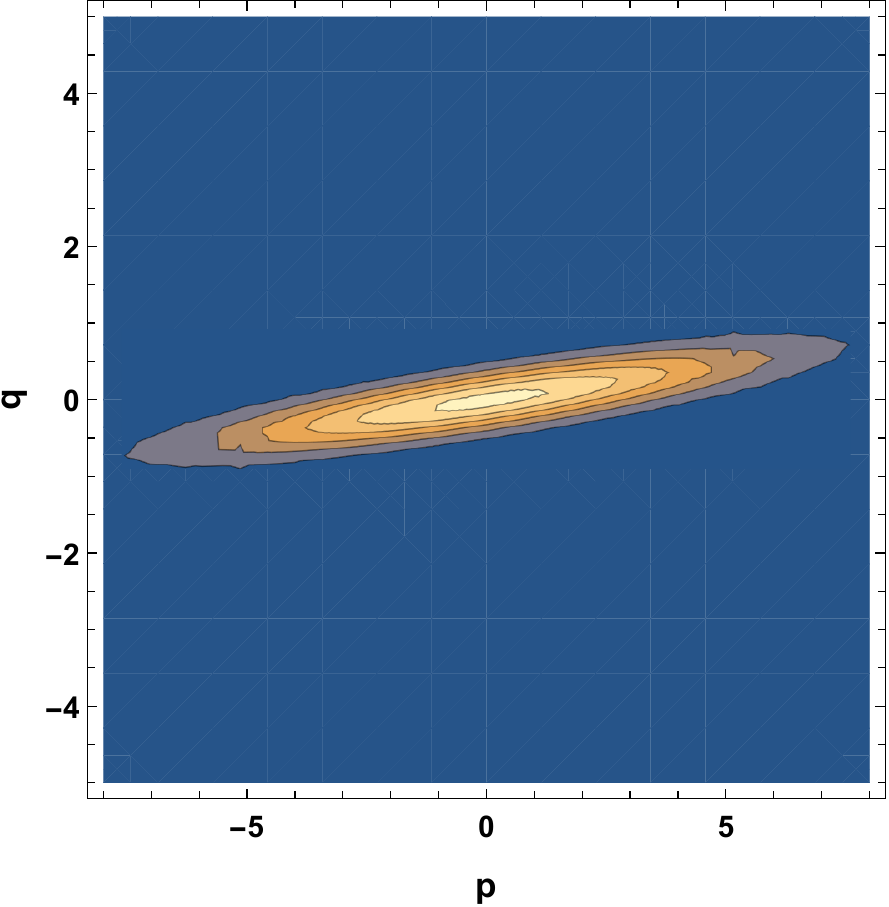} & 
		\includegraphics[width=0.33\textwidth]{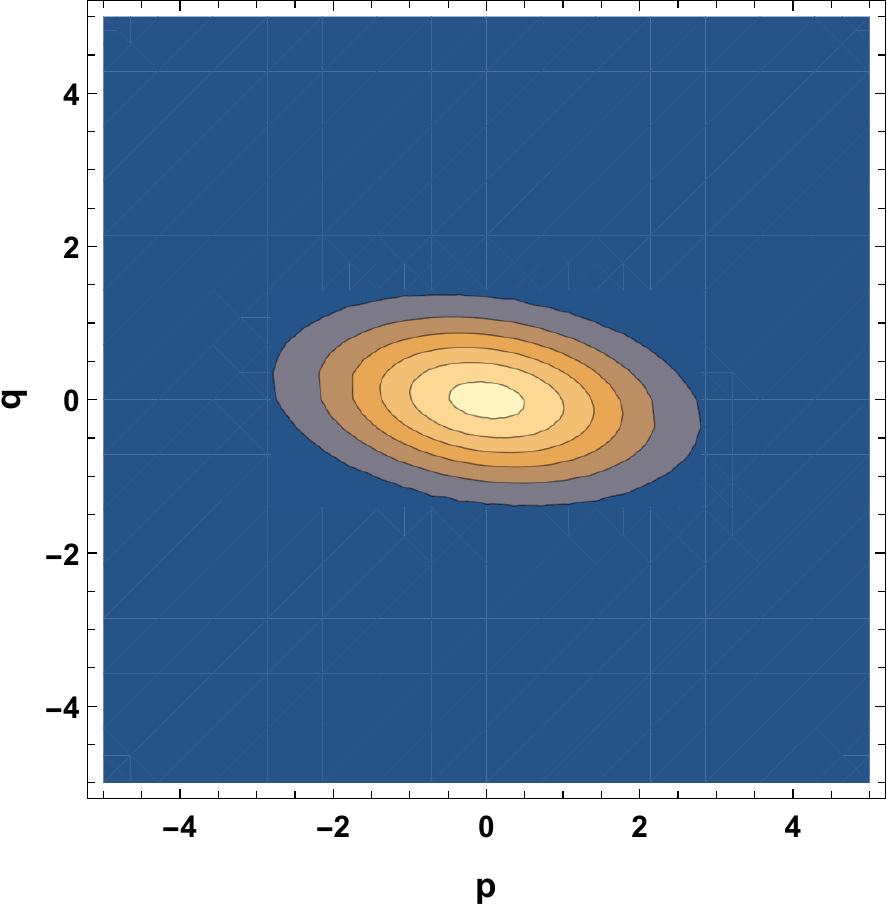}  \\

        (a) & (b) & (c) \\
	                
	\end{tabular}
\includegraphics[width=0.09\textwidth, angle=270]{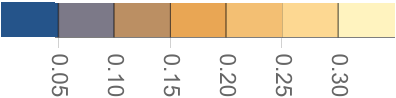}	

\caption{Wigner function in stable zone at  $t=0$ in (a), $t=3.5$ in (b), $t=7$ in (c). The values of the parameters used are $g=4.0$, $h=2$ and $\gamma=0.5$. }
\label{fig:wigner_stable}
\end{figure}

\begin{figure}[h]
	\centering
	\begin{tabular}{ccc}
		\includegraphics[width=0.33\textwidth]{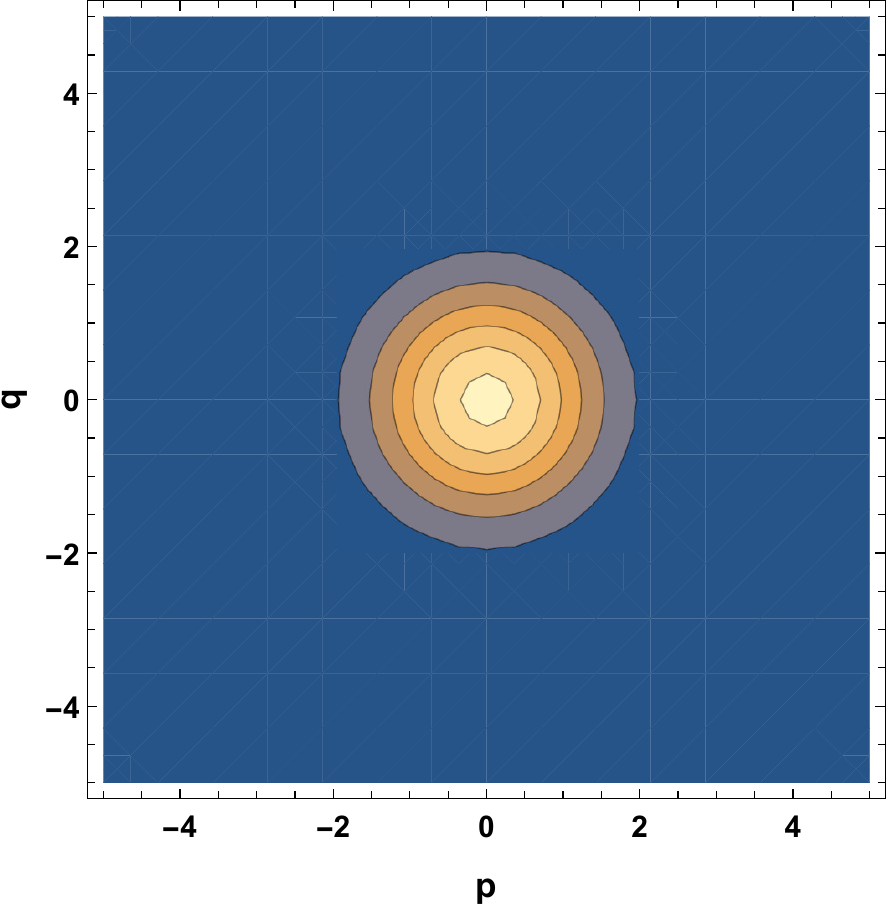} & 
		\includegraphics[width=0.33\textwidth]{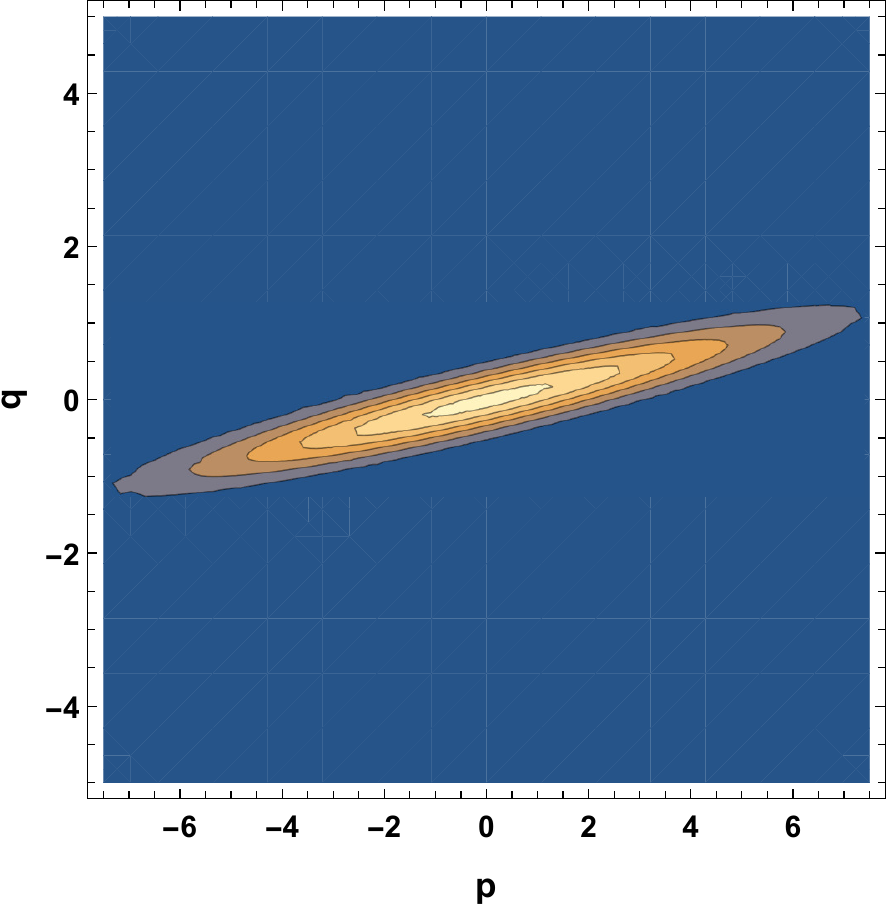} & 
		\includegraphics[width=0.33\textwidth]{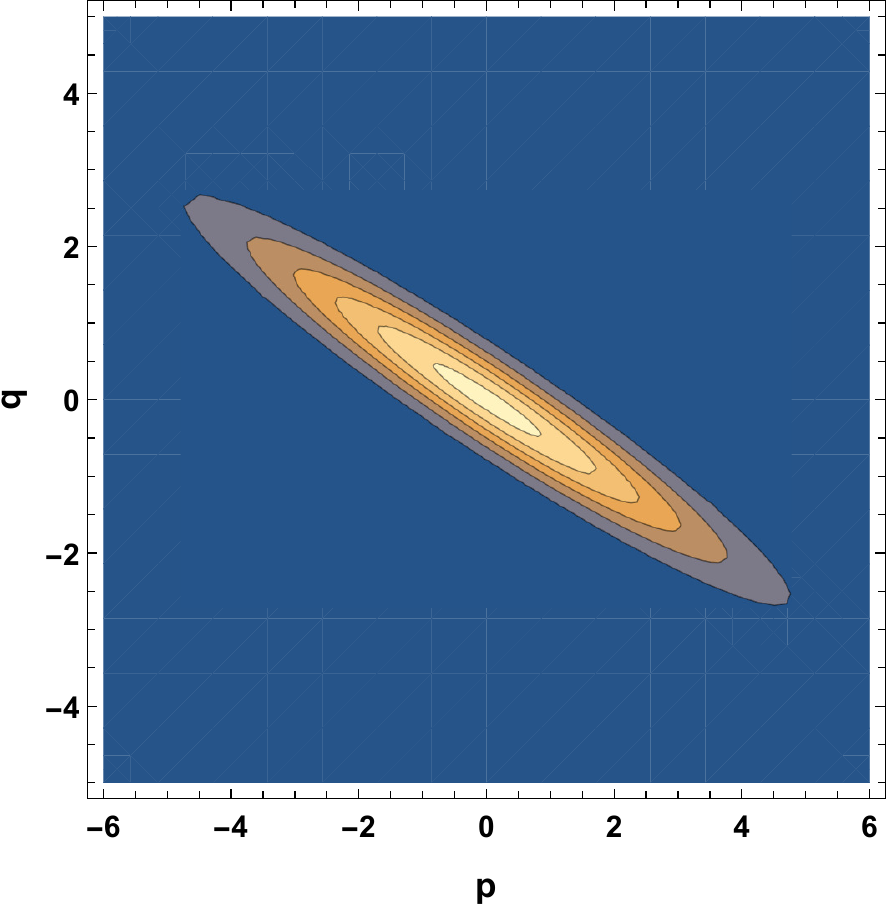}  \\

        (a) & (b) & (c) \\
	                
	\end{tabular}
\includegraphics[width=0.09\textwidth, angle=270]{color_bar.pdf}	

\caption{Wigner function in unstable zone at  $t=0$ in (a), $t=3.5$ in (b), $t=7$ in (c). The values of the parameters used are $g=4.5$, $h=2.5$ and $\gamma=0.5$. }
\label{fig:wigner_unstable}
\end{figure}

\noindent To get a deeper understanding of the physics underlying this dissipative system, we consider now the problem from another point of view: the Wigner  function representation. It is defined by the position representation of the density matrix \cite{tannor2007introduction}, 
\begin{equation}
\begin{split}
W(q,p,t)&=\frac{1}{2\pi}\int_{-\infty}^\infty dq  e^{i p s} \left< q-\frac{s}{2}| \hat{\rho} |  q+\frac{s}{2} \right>.  \\
%&= \frac{1}{2\pi}\int_{-\infty}^\infty dq  e^{i p s}  \hat{\rho} \left(q+\frac{s}{2}, q-\frac{s}{2}, t \right)
\label{eq:ground_wigner}
\end{split}
\end{equation}
\noindent Where the density matrix is defined as $\hat{\rho}=|0,t\rangle \langle 0,t|$ and in the coordinate representation, it will be, 
\begin{equation}
\begin{split}
\hat{\rho}(x,x_1, t) &= \psi_0(x,t) \psi^*_0(x_1,t) \\
&=\left(\frac{1}{2 \pi |B(t)|^2}\right)^\frac{1}{2} exp\left( \frac{i \dot{B}(t)e^{\gamma t}}{2 B(t)}x^2\right) exp\left(- \frac{ i \dot{B^*}(t)e^{\gamma t}}{2 B^*(t)}x_1^2\right).
\label{eq:ground_density_matrix}
\end{split}
\end{equation}
\noindent The Wigner function can take on negative as well as positive values. However, for the pure state, the Wigner function is Gaussian and hence is positive definite. Because of the positivity of the Wigner function, it can be interpreted as a ``phase space distribution", similar to a probability distribution of classical particles. %Using Fourier transformation $\left(f(\omega)=\frac{1}{\sqrt{2 \pi}} \int_{-\infty}^\infty dt f(t) e^{i \omega t}\right)$,
Using Eq. (\ref{eq:ground_density_matrix}), we get the following results, 
\begin{equation}
\begin{split}
W(q,p,t) %&= \frac{e^{- \frac{q^2}{2 |B(t)|^2}}}{2 \pi \sqrt{ 2 \pi |B(t)|^2}} \int_{-\infty}^\infty ds e^{i \left[p-\frac{e^{\gamma t}}{2} \left(\frac{\dot{B}(t)}{ B(t)}+\frac{\dot{B}^*(t)}{ B^*(t)}\right)q\right]s } e^{-\frac{s^2}{8 |B(t)|^2}} \\
%&= \frac{e^{- \frac{q^2}{2 |B(t)|^2}}}{2 \pi \sqrt{ |B(t)|^2}} \frac{}{\sqrt{2} \sqrt{\frac{1}{8 |B(t)|^2}}}\\
&=\frac{1}{\pi} \exp\left[{- \frac{q^2}{2 |B(t)|^2}} -\frac{\left[p-\frac{e^{\gamma t}}{2} \left(\frac{\dot{B}(t)}{ B(t)}+\frac{\dot{B}^*(t)}{ B^*(t)}\right)q\right]^2 |B(t)|^2}{2}\right].
\label{eq:ground_wigner_solution}
\end{split}
\end{equation}
\noindent However, the term, $\left(\frac{\dot{B}(t)}{ B(t)}+\frac{\dot{B}^*(t)}{ B^*(t)}\right)$ is a pure real number. In Fig.  $(\ref{fig:wigner_stable})$ and $(\ref{fig:wigner_unstable})$, we plot the evolution of $W(q,p,t) $ at three different time intervals for the stable (blue diamond) and unstable (red circle) points respectively. In both the stable and unstable zone, $W(q,p,t) $ evolves in such a way that it is stretched along a dynamical axis with a consequent contraction along it's perpendicular axis in the phase space. This is happening because of the time-dependent nature of uncertainty principle. In comparison, the behaviour of $W(q,p,t) $ for the stable zone is different from the unstable zone. For the unstable zone, the stretching of the $W(q,p,t) $ is observed for all time (Fig. $\ref{fig:wigner_unstable}$ (a)-(c)). However, for the stable zone, $W(q,p,t) $ stretches along the diagonal axis at an intermediate time (Fig.  $\ref{fig:wigner_stable} (b))$ and revives its shape at the final time  (Fig.  $\ref{fig:wigner_stable} (c))$.  Hence, the overall squeezing of $W(q,p,t) $ in stable zone is less  compared to that in the unstable zone.

\section{Conclusion}
\noindent In this work, we have studied an effective Hamiltonian dynamics of a single quantum harmonic oscillator in the presence of dissipation and parametric driving. The time-dependent parametric frequency drives the system perpetually out of equilibrium and the stable points of the undriven system become unstable or an unstable point may become a stable one. But the role of dissipation is always uni-directional which means it always pushes the system from an unstable zone to a stable one. By tuning the parameters of the system, one can have control over the competition between dissipation and parametric driving. In addition, we have studied the relationship between the dissipation and the localization of the system’s wave function in the Wigner function representation. Dissipation modifies the functional form of the Wigner function. In the phase space, the de-localization of the Wigner function decreases with the dissipation because dissipation stabilizes the system. Hence, a direct correspondence between the dynamical stability of the time-dependent parameters and the temporal behaviors of the Wigner function is also observed in an isolated quantum damped harmonic oscillator system with parametric driving.

\begin{acknowledgments}
SC would like to thank Dr. Ashwani Tripathi for the help to learn MATHEMATICA.  SC acknowledges DST-Inspire scholarship for the financial help during his M.Sc.
\end{acknowledgments}

\end{document}